\documentclass[preprint,showpacs,prb,floatfix,superscriptaddress,citeautoscript
,cite]{revtex4}
\usepackage{graphicx}
\usepackage{epstopdf}
\usepackage{color}
\usepackage{amsmath}

\begin{document}

\title{Electronic and Vibrational Properties of PbI$_2$: From Bulk to Monolayer}

\author{M. Yagmurcukardes}
\email{Mehmet.Yagmurcukardes@uantwerpen.be}
\affiliation{Department of Physics, University of Antwerp, Groenenborgerlaan
171, B-2020 Antwerp, Belgium}

\author{F. M. Peeters}
\affiliation{Department of Physics, University of Antwerp, Groenenborgerlaan
171, B-2020 Antwerp, Belgium}

\author{H. Sahin}
\affiliation{ICTP-ECAR Eurasian Center for Advanced Research, Izmir Institute 
of 
Technology, 35430, Izmir, Turkey}
\affiliation{Department of Photonics, Izmir Institute of Technology, 35430, 
Izmir, Turkey}

\date{\today}

\begin{abstract}

Using first-principles calculations, we study the dependence of the  
electronic 
and vibrational properties of multi-layered PbI$_2$ crystals on the number of 
layers and focus on the 
electronic-band structure and the Raman spectrum. Electronic-band structure 
calculations 
reveal that the direct or indirect 
semiconducting behavior of PbI$_2$ is strongly influenced by the number of 
layers. 
We find that at 3L-thickness there is a direct-to-indirect band gap 
transition (from bulk-to-monolayer). It is shown that in the Raman 
spectrum two prominent peaks, A$_{1g}$ and E$_g$, 
exhibit phonon hardening with increasing number of layers due to the 
inter-layer van der Waals interaction. Moreover, the Raman activity of the 
A$_{1g}$ 
mode 
significantly 
increases with increasing number of layers due to the enhanced out-of-plane 
dielectric constant in the few-layer case. We further 
characterize 
rigid-layer vibrations of low-frequency inter-layer
shear (C) and breathing (LB) modes in few-layer PbI$_2$. A reduced 
mono-atomic (linear) chain model (LCM) provides a
fairly accurate picture of the number of layers dependence of the low-frequency
modes and it is shown also to be a powerful tool to study the
inter-layer coupling strength in layered PbI$_2$.

\end{abstract}

\maketitle
\section{Introduction}
Over the past decade, successful synthesis of graphene\cite{Novo1,Geim1} 
led to an enormous interest in the field of two
dimensional (2D) materials. However, the lack of a band gap in graphene 
restricted its applications and search for other 2D materials with a suitable 
band gap became necessary. With this 
respect, 
many other 2D monolayer materials such as silicene,\cite{Cahangirov,Kara} 
germanene\cite{Cahangirov}, group III-V binary 
compounds ($h$-BN, $h$-AlN)\cite{Sahin3,Wang2,Kim,Tsipas,Bacaksiz} and 
transition-metal dichalcogenides 
(TMDs)\cite{Gordon,Coleman,Wang1,Ross,Sahin2,Tongay,Horzum,Chen3} 
 were successfully synthesized. Recently a post-transition metal iodide, 
PbI$_2$, was added to the library of 2D monolayer 
materials\cite{Zhong}. 

Lead iodide (PbI$_2$) is a typical layered van der Waals (vdW) crystal in its 
bulk 
form which crystallizes in the well-known 1T phase. The PbI$_2$ units are also 
known to form lead halide perovskites which were recently 
investigated.\cite{h1,h2}  
Its bulk crystal is 
composed 
of covalently 
bonded I-Pb-I
repeating layers that interact weakly with vdW 
forces\cite{Toulouse,Baibarac,Preda,Zhang,Liu}. The bulk crystal of PbI$_2$ was 
demonstrated to be a good semiconductor for photoluminescence, 
electroluminescence, and non‐linear
optical field applications\cite{Guloy,Cabana}. In addition, thickness-dependent 
optoelectronic
properties of PbI$_2$ is another important feature of the material. Toulouse 
\textit{et al.} found theoretically that the electronic-band structure 
of PbI$_2$ exhibits a shift from direct-gap with 2.38 eV to an indirect-gap 
semiconductor
with 2.5 eV when its thickness is thinned down to
bilayer or monolayer\cite{Toulouse}. In another study, Zhou \textit{et al.} 
investigated the structural stability and strain-dependent electronic 
properties of monolayer PbI$_2$ and showed that the band gap of the material is 
tunable under biaxial strain in a wide energy range of 1-3 eV\cite{Zhou}. Wang 
\textit{et al.} confirmed experimentally the thickness- and 
strain-dependent 
photoluminescence properties of PbI$_2$\cite{Wang} and reported that 
thickness-dependent vdW epitaxial strain can be significant and influences 
substantially the
photoluminescence properties of PbI$_2$. Very recently, Zhong \textit{et al.} 
successfully synthesized large scale monolayer and few-layer PbI$_2$ with 
high crystallinity using the physical vapor deposition (PVD) method\cite{Zhong} 
and using photoluminescence measurements 
showed direct-gap to indirect-gap transition in PbI$_2$ when going 
from bulk to monolayer.

One of the most common technique 
for the characterization of a material is Raman spectroscopy\cite{raman} which 
gives information about the structural phase of the material by monitoring the 
characteristic vibrational energy 
levels of the sample. Raman measurement can give information about 
the substrate-free 
number of layers identification of layered 
materials,\cite{ferrari,qiao,xzhang} the 
strength of the inter-layer coupling in layered materials\cite{xzhang,phtan} 
and 
interface coupling in vdW 
heterostructures.\cite{jbwu,jbwu2} Absolute and relative activities of the 
Raman peaks lead to 
the determination of different phase distributions in a 
material.\cite{colom,gouadec,havel}
Raman spectroscopy can also give information about the electronic structure, 
thickness, and can be used to probe strain, stability, stoichiometry, and 
stacking orders of 2D 
materials.\cite{xzhang2}

The PbI$_2$ crystal is known as a good semiconductor for photoluminescence, 
electroluminescence, and non‐linear
optical field applications which is also known to possess important 
thickness-dependent 
optoelectronic
properties. The thickness-dependent electronic properties of PbI$_2$ were 
already 
investigated by means of photoluminescence measurements and \textit{ab-initio} 
calculations. Here, we aim to study the number of layer dependency of 
the electronic-band structure of PbI$_2$ and explain the physical origin of 
the indirect-to-direct band gap transition. In addition, we investigate, for 
the 
first 
time, 
the 
layer-dependent vibrational 
properties of PbI$_2$ in terms of high-frequency prominent optical peaks and 
low-frequency 
inter-layer shear (C) and breathing (LB) modes in order to get 
information 
about the layer-layer interaction in few-layer PbI$_2$. 

The paper is organized as 
follows: Details of the computational methodology and 
Raman scattering theory are 
given in Sec. \ref{comp}. The evolution of electronic-band structure with the 
number of layers is discussed in Sec. \ref{electronic}. In Secs. 
\ref{hf} and \ref{lf}, the evolution of the Raman spectrum of PbI$_2$ from bulk 
to 
monolayer is discussed in terms of the peak frequencies and Raman 
activities of high-frequency optical and low frequency inter-layer shear and 
breathing modes, respectively.

\section{Computational Methodology}\label{comp}

To investigate the structural, electronic and vibrational properties of PbI$_2$ 
crystals, 
first principle calculations 
were performed in the framework of density functional theory (DFT) as 
implemented 
in the Vienna \textit{ab-initio} 
simulation package (VASP).\cite{vasp1,vasp2} The Perdew-Burke-Ernzerhof 
(PBE)\cite{perdew} form of generalized gradient 
approximation (GGA) was adopted to describe electron exchange and correlation. 
The van der Waals (vdW) correction to the GGA 
functional was included by using the DFT-D2 method of Grimme.\cite{grimme} The 
electronic-band structures were calculated 
with the 
inclusion of spin-orbit-coupling (SOC) on top of GGA and 
Heyd-Scuseria-Ernzerhof 
(HSE)\cite{hse} 
screened-nonlocal-exchange functional of the generalized Kohn-Sham scheme, 
respectively. The 
charge transfer in the system was determined by 
the Bader technique.\cite{bader} 

The kinetic energy cut-off for plane-wave 
expansion was set to 500 eV and the energy was minimized until its variation in 
the following steps became less than 10$^{-8}$ eV.
The Gaussian smearing method was employed for the total energy calculations. 
The 
width of the smearing was chosen to be 
0.05 eV. Total Hellmann-Feynman forces was taken to be 10$^{-7}$ 
eV/\AA {} for the structural 
optimization. 18$\times$18$\times$1 $\Gamma$ centered \textit{k}-point 
samplings 
were used in the primitive unit cells. To avoid interaction between the 
neighboring 
layers, 
the calculations
were implemented with a vacuum space of 25 \AA {}.

The phononic properties of PbI$_2$ crystals were calculated in 
terms 
of the off-resonant Raman activities of the phonon modes at the $\Gamma$ point. 
For 
this purpose, the zone-centered vibrational phonon modes were 
calculated using the finite-difference method as implemented in VASP. Each atom 
in 
the primitive unit cell was initially distorted by 0.01 \AA {} and the 
corresponding dynamical matrix was constructed. Then, the vibrational modes 
were determined by a direct diagonalization of the dynamical matrix. The 
kinetic 
energy cut-off for plane-wave 
expansion was increased to 800 eV with a $k$-point set of 
24$\times$24$\times$1 in the case of Raman calculations. 
The $k$-point set and kinetic energy cut-off were systematically increased step 
by step until convergence for the 
frequencies of acoustic modes was reached (0.0 cm$^{-1}$ for each acoustic 
mode at the $\Gamma$ 
point). Once the accurate phonon mode frequencies were obtained at the $\Gamma$ 
point, the change of the macroscopic dielectric tensor was calculated with 
respect 
to each 
vibrational 
mode to get the corresponding Raman activities\cite{vasp-raman}.

In a 
Raman scattering experiment, the sample is exposed to light and instantly 
scattered 
photons 
are collected. The dispersion of the collected photons with respect to a shift 
in frequency 
gives the Raman spectrum. In Raman theory, the inelastically scattered 
photon originates from the oscillating dipoles of the crystal corresponding 
to the Raman active vibrational modes.

The treatment of Raman activities is based on Placzek's classical
theory of polarizability\cite{Placzek}. According to the Placzek approximation, 
the 
activity of a Raman active phonon mode is 
proportional to $|\hat{e}_s.R.\hat{e}_i|^2$ where $\hat{e}_s$ and $\hat{e}_i$ 
stand for the polarization 
vectors of scattered radiation and incident light, respectively. $R$ is a 
3$\times$3, second rank tensor known as the Raman tensor 
whose elements are derivatives of polarizability of the material with respect 
to the 
phonon 
normal modes,  

\begin{equation}
R=\left[\begin{array}{ccc}
    \frac{
\partial\alpha_{11}}{\partial 
Q_k} & \frac{
\partial\alpha_{12}}{\partial 
Q_k} & \frac{
\partial\alpha_{13}}{\partial 
Q_k} 	\\
    \frac{
\partial\alpha_{21}}{\partial 
Q_k} & \frac{
\partial\alpha_{22}}{\partial 
Q_k} & \frac{
\partial\alpha_{23}}{\partial 
Q_k} 	\\
    \frac{
\partial\alpha_{31}}{\partial 
Q_k} & \frac{
\partial\alpha_{32}}{\partial 
Q_k} & \frac{
\partial\alpha_{33}}{\partial 
Q_k} 	\\
\end{array}\right]
\end{equation}where $Q_k$ is the normal mode describing the whole motion of 
individual 
atoms participating to the $k^{th}$ vibrational phonon mode while $\alpha_{ij}$ 
is the 
polarizability tensor of the material. The term $|\hat{e}_s.R.\hat{e}_i|^2$ is 
called the 
Raman activity which is calculated from the 
change of polarizability. For a back scattering experimental geometry the total 
Raman activity 
is represented in terms of Raman invariants given by,

\begin{align}\label{invariants} 
\tilde{\alpha}_s \equiv &\frac{1}{3} 
(\tilde{\alpha}_{xx}+\tilde{\alpha}_{yy}+\tilde{\alpha}_ {zz}), \\
  \beta \equiv &\frac{1}{2} 
\{(\tilde{\alpha}_{xx}-\tilde{\alpha}_{yy})^2+(\tilde{\alpha}_{yy}-\tilde{\alpha
}
_{zz})^2+(\tilde{\alpha}_{zz}-\tilde{\alpha}_{xx})^2 \nonumber \\ 
&+6[(\tilde{\alpha}_{xy})^2+(\tilde{\alpha}_{yz})^2+(\tilde{\alpha}_{xz})^2]\}, 
\end{align}where $\tilde{\alpha}_s$ and $\beta$ represent the isotropic and 
anisotropic 
parts of  
the derivative of the polarizability tensor with respect to the phonon normal 
mode, 
respectively. The importance of such representation is its invariance 
under a change in the sample orientation. 
Finally, using these forms of symmetric and anti-symmetric polarizability 
derivative tensors, the Raman activity, $R_A$, can be written as,

\begin{equation}\label{activity-final} 
R_A=45\tilde{\alpha}_s^2+7\beta^2.
\end{equation}In the rest of the paper, the Raman activities of PbI$_2$ 
crystals 
are calculated using Eq. (\ref{activity-final}).

\begin{figure}
\includegraphics[width=8cm]{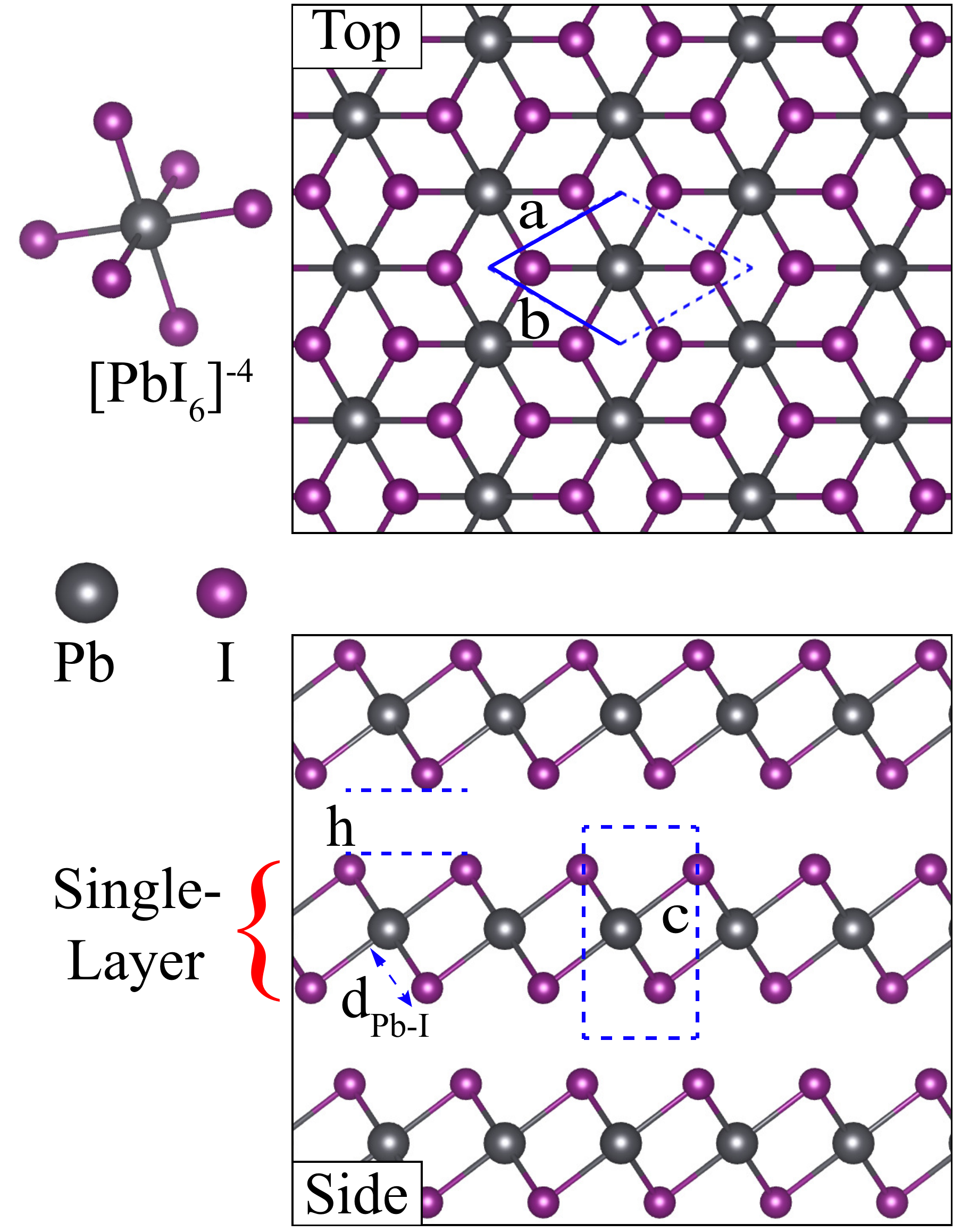}
\caption{\label{f1}
(color line) Top and side views of bulk PbI$_2$. The lattice parameters 
$a$, $b$, and $c$, the octahedral unit, [PbI$_6$]$^{-4}$, layer thickness, and 
Pb-I atomic bond length, d$_{Pb-I}$, are shown in the inset of the figures. For 
the visualization of the atomic structure the software VESTA was 
used.\cite{vesta} }
\end{figure}

\begin{table}
\caption{\label{main} From bulk to monolayer PbI$_2$ crystal, the thickness of 
PbI$_2$ layers, h, 
energy band gaps including SOC, E$_\textrm{gap}^{SOC}$, and HSE06, 
E$_\textrm{gap}^{HSE06+SOC}$. Location of VBM and CBM edges in the BZ, and the 
work 
function, $\Phi$     
.}
\begin{tabular}{rcccccccccccccccc}
\hline\hline
& & &  & & & & 
  
\\
&h&E$_\textrm{gap}^{SOC}$& 
E$_\textrm{gap}^{SOC+HSE06}$& VBM/CBM &$\Phi$ \\
& (\AA{})& (eV)&(eV)&$-$&(eV)\\
\hline
1L-PbI$_2$&7.13  &1.99 &2.65 & M$-$$\Gamma$/$\Gamma$&6.10\\
2L-PbI$_2$ &21.39 &1.75 &2.39& M$-$$\Gamma$/$\Gamma$&5.99\\
3L-PbI$_2$&28.52 & 1.62 &2.23 & $\Gamma$/$\Gamma$&5.93\\
4L-PbI$_2$ &35.65 & 1.50 &2.18& $\Gamma$/$\Gamma$& 5.84\\
5L-PbI$_2$& 42.78&1.47 &2.14 &$\Gamma$/$\Gamma$& 5.82\\
6L-PbI$_2$&49.91  &1.45 &2.11(2.38)\cite{Zhong} & $\Gamma$/$\Gamma$&5.80\\
Bulk-PbI$_2$& $-$  &1.40&2.07(2.41)\cite{Zhong}& A/A&$-$\\
\hline\hline 
\end{tabular}
\end{table}

\section{Monolayer-\lowercase{to}-Bulk P\lowercase{b}I$_2$}

\begin{figure*}
\includegraphics[width=18cm]{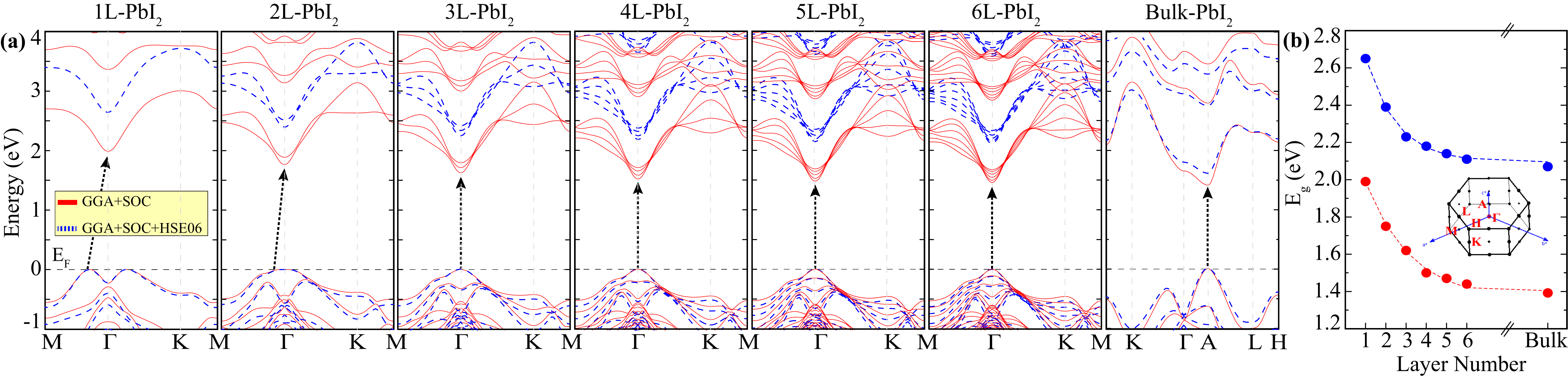}
\caption{\label{f2}
(color line) (a) Evolution of electronic-band structure 
from monolayer to bulk crystal of PbI$_2$. The Fermi 
energy 
($E_F$) level  is
set to the valence band maximum. The red solid and blue dashed lines represent 
the band structures calculated within SOC and HSE06 on top of 
GGA, respectively. (b) The change of band gap with 
respect to the number of layers. The inset shows the high symmetry points in 
the BZ. }
\end{figure*}

\begin{figure}
\includegraphics[width=8.5cm]{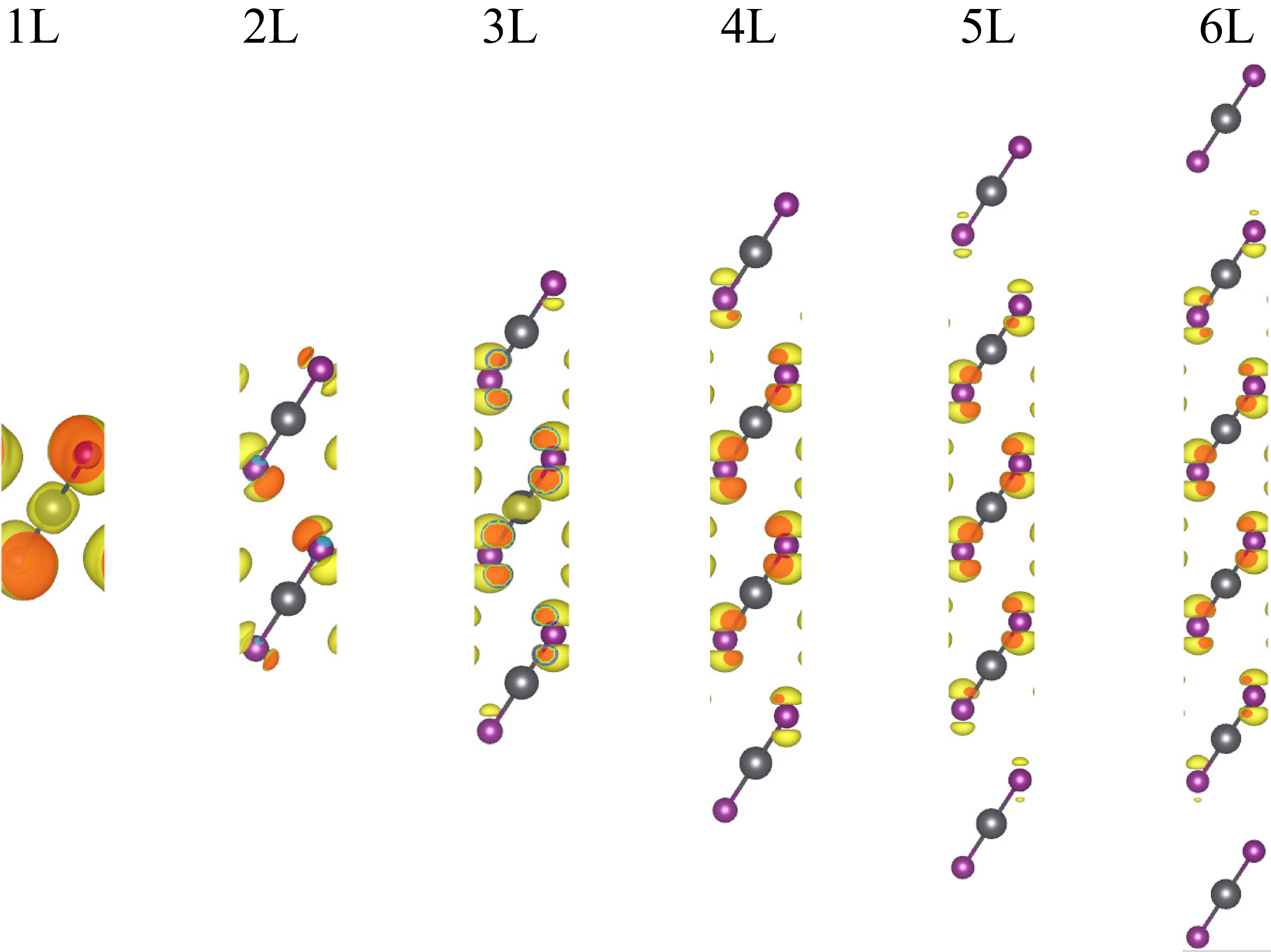}
\caption{\label{f3}
(color line) Atomic orbital character of the valence band maximum of PbI$_2$ 
crystals from 1L-to-6L. The isosurface value is 
5$\times$10$^{-6}$ $e$/\AA$^3$ {}. The atomic orbitals were visualized using 
the software VESTA.\cite{vesta} }
\end{figure}

\subsection{Electronic-Band Structure}\label{electronic}
Like many TMDs, monolayer PbI$_2$ 
crystallizes in 
either the 1H or 1T phase. It was already 
demonstrated that the 1T phase is the structural ground 
state of monolayer PbI$_2$ \cite{cihan}. In the present study, 
we consider the 1T phase for PbI$_2$ crystal (see Fig. \ref{f1}). 1T phase of 
bulk- and 
monolayer-PbI$_2$ can be represented by a 3-atom primitive unit cell. The bulk 
crystal is composed of weakly vdW interacting PbI$_2$ 
layers. In an isolated 
monolayer crystal, 
a layer of Pb atoms is sandwiched between two layers of 
I atoms which corresponds to the space group $P\bar{3}m2$. Each Pb atom is 
surrounded by 6 I-atoms forming a near-octahedral 
[PbI$_6$]$^{4-}$ unit. When sharing edges with six neighboring octahedra a 
monolayer of PbI$_2$ is constituted (see left panel of Fig. \ref{f1}). The 
calculated in-plane and 
out-of-plane 
lattice parameters for the bulk crystal are 4.45 and 7.09 \AA {}, respectively. 
The Pb-I atomic bond length is 3.23 \AA {} while the inter-layer distance is 
3.18 \AA {}. In the 
case of few-layer PbI$_2$ crystals, 
the in-plane lattice constant slightly decreases (4.44 \AA 
{}) with the 
corresponding Pb-I bond length of 3.24 \AA {}. Thus, it is important to note 
that the structural parameters are almost independent on the number of layers 
of 
PbI$_2$. Bader charge
analysis shows that an amount of $\sim$0.4 $e^-$ is received by an I atom 
indicating the ionic bonding character between Pb and I atoms. In addition, as 
listed in Table \ref{main}, the work function ($\Phi$) which is defined for a 
semiconductor as the amount of energy required to remove a charge carrier 
located at 
the Fermi energy to vacuum as a free particle, decreases rapidly from 
monolayer to 4L-crystal and then slowly upon further increasing 
the number of layers. The reason for such decrease is that as the number of 
layers 
increases, the number of electrons also increases which sets the Fermi level to 
higher energies. This leads to a decrease in work function which is the energy 
difference between the vacuum level and the Fermi level. 

In order to understand the effect of the thickness on the electronic properties 
of 
PbI$_2$ crystals, we 
perform electronic-band structure calculations for different thicknesses of 
PbI$_2$ 
crystals (1L, 2L, 3L, 4L, 5L, 6L, and bulk). As shown in Fig. \ref{f2}(a), 
the conduction band minimum (CBM) is located at the $\Gamma$ 
point in the BZ for all PbI$_2$ crystals. However, as the number of layers 
increases from 1L to 3L, the valence band maximum (VBM) shifts from between the 
$\Gamma$ and K points to the $\Gamma$ point which 
indicates a transition from indirect-to-direct band gap for 3L-PbI$_2$. In this 
section, we give our HSE06+SOC band gap results which approximately gives the 
correct 
band gap for PbI$_2$.
The indirect band gap values are 2.65 eV and 2.39 eV for 
1L and 2L 
crystals, 
respectively. For 3L and thicker structures, 
the VBM shifts to the $\Gamma$ point and the direct 
band gap for 
3L-PbI$_2$ 
is 2.23 eV. The thickness of 
3L-PbI$_2$ ($\sim$2.9 nm) seems to be the critical thickness for 
such an indirect-to-direct band gap transition. As the number of layer 
increases 
to 
6L, the band gap decreases to 2.11 eV and saturates to 2.07 eV 
for 
bulk-PbI$_2$. Our results for the direct-to-indirect band gap 
transition agree with those reported by Toulouse \textit{et 
al.}\cite{Toulouse}. However, quantitative differences between our 
band gap results and theirs are due to the use of different functionals. 
Different 
from the methodology used by Toulouse \textit{et 
al.}\cite{Toulouse}, we consider the GGA functional within vdW correction which 
is very important for layered materials. Although, it was pointed out by 
Toulouse \textit{et 
al.}, the nature of the band gap transition in PbI$_2$ is explained through 
layer-layer interaction while the change in the band gap is driven by 
both quantum confinement and vdW inter-layer interaction. In 
addition, we aim to understand the behavior of the band gap with the number of 
layers 
by fitting the band gap values to a functional of the form given in Eq. 
(\ref{fitting}). 

In order to compare with the usual particle in a box model for quantum 
confinement for which the energy 
decays as $\sim$1/$N^2$\cite{confinement}, we fitted the band gap to 
a general power law of the form\cite{fit3}:
\begin{equation}\label{fitting} 
E_{gap}(N)=E_{gap}(bulk)+\frac{A}{N^{\kappa}},
\end{equation}where $N$ is the number of layers. The value 
E$_{gap}$(bulk), 2.07 eV, is the bulk band gap and we obtain $\kappa$ and $A$ 
to 
be 1.3 and 0.6 eV, respectively. 
Here since two 
different physical mechanisms drive the changes in the band structures, one 
may also try to fit the band gap change to a function including both 
exponential 
and power law forms as suggested by Rudenko \textit{et al.}\cite{fit2}. In 
addition, Tran \textit{et al.}\cite{fit3} demonstrated that the quantum 
confinement exponent may give different values for the band gap fits calculated at different levels of the theory (PBE, GW,...).
Indeed, it should be noted that the change in the band gap is different from 
1L-to-3L and from 4L-to-bulk. In the first thickness regime, the 
orbital-orbital interaction between neighboring PbI$_2$ layers dominate the vdW 
interaction and drives the indirect-direct band gap transition. However, for 
thicker crystals the electrons are mostly confined to the layers and the 
relatively small decrease in the band gap can be attributed to weak 
interactions 
between the layers (vdW, Coulomb etc...) 

The transition from 
indirect-to-direct band gap semiconducting 
behavior can be attributed to the orbital 
hybridizations between I atoms from the nearest neighboring layers. As shown in 
Fig. 
\ref{f3}, in monolayer PbI$_2$ the VBM is composed of 
mixed in-plane $p$-orbitals ($p_x$ and $p_y$). When the second layer is 
introduced, i.e. the bilayer case, 
the VBM is composed of tilted-interacting $p_z$-orbitals of the I 
atoms. As the number of layers increases to 3, the hybridization between 
the I atoms from neighboring layers converts the VBM orbitals completely to 
$p_z$-orbitals 
which controls the indirect-to-direct band gap semiconducting transition. In 
few-layer PbI$_2$ crystals, it is seen that this 
hybridization 
mostly occurs between the I atoms of the internal layers and thus the 
contributions from the outer layers become negligible. In contrast to the VBM, 
the 
CBM consists of $p$-orbitals of the Pb atoms which are located 
at the center of each layer. The CBM has no thickness dependency 
since there is no direct interaction between the Pb atoms of the neighboring 
layers.

\begin{figure*}
\includegraphics[width=18cm]{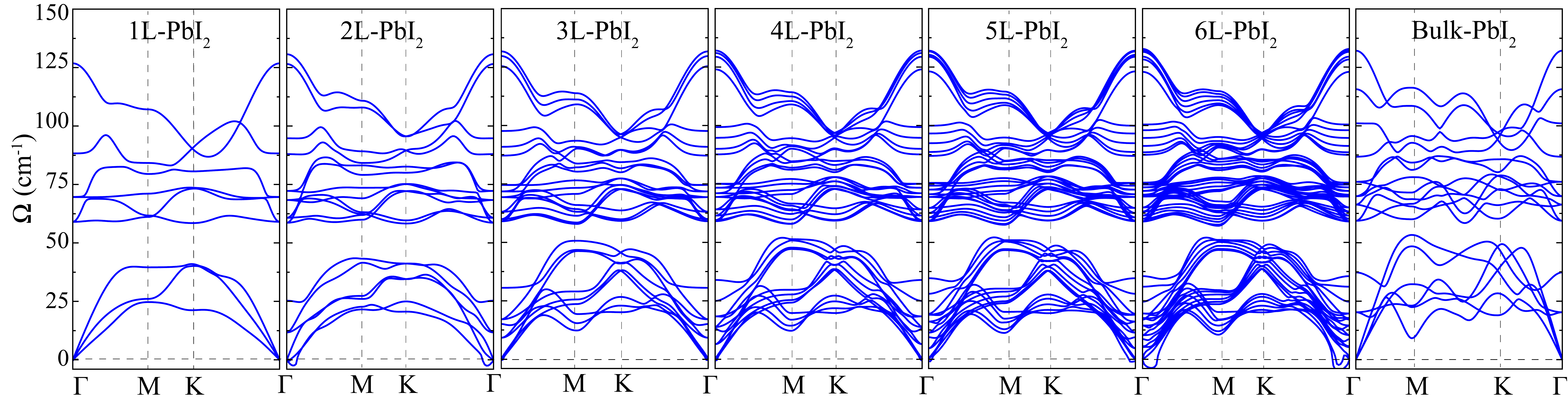}
\caption{\label{f4}
(color line) The phonon-band structures of PbI$_2$ crystals from monolayer 
to bulk. }
\end{figure*}

\begin{table*}
\caption{\label{main2} From bulk to monolayer PbI$_2$: calculated and 
fitted peak 
positions 
of 
the Raman active phonon modes, E$_g$(calc), E$_g$(fit), A$_{1g}$(calc), and 
A$_{1g}$(fit), the relative frequency shift of 
each phonon mode with respect to the frequency of the mode in ($N$-1)-PbI$_2$, 
$\bigtriangleup\Omega_{A_{1g}}$ and 
$\bigtriangleup\Omega_{E_{g}}$. The Raman activity of phonon modes and their 
relative ratios, 
I$_{E_g}$, I$_{A_{1g}}$, and $\frac{I_{A_{1g}}}{I_{E_g}}$. The in-plane 
($\epsilon_{in}$) and out-of-plane ($\epsilon_{out}$) static (low-frequency) 
dielectric 
constant. The 
frequencies given in the parentheses are from the literature.   }
\begin{tabular}{rcccccccccccccccc}
\hline\hline
& & &  & & & & 
  
\\
& E$_g$(calc) & E$_g$(fit) & A$_{1g}$(calc) & A$_{1g}$(fit) & 
$\bigtriangleup\Omega_{A_{1g}}$ & $\bigtriangleup\Omega_{E_{g}}$ &I$_{E_g}$ 
&I$_{A_{1g}}$ 
&$\frac{I_{A_{1g}}}{I_{E_g}}$ & 
$\epsilon_{in}$ 
&$\epsilon_{out}$\\
& (cm$^{-1}$) & 
(cm$^{-1}$) & (cm$^{-1}$) & 
(cm$^{-1}$) &(\%) &(\%) & ($\frac{\AA{}^4}{amu}$) & ($\frac{\AA{}^4}{amu}$) & 
$-$ & $-$& 
$-$  \\
\hline
1L-PbI$_2$&68.5  & $-$  & 87.6  & $-$  &$-$ &$-$ 
&1.3 &0.1 &0.1 &2.56 & 1.30 \\
2L-PbI$_2$&73.1  & 73.0 & 94.5(96.0)\cite{Zhong}  & 94.4  &7.9 &6.7 
&2.0 &3.2 &1.6 & 3.24 & 1.49 \\
3L-PbI$_2$& 74.4  & 74.5& 97.7  & 97.9&3.4 &1.9
&3.3 &14.3 &4.3 & 3.84 & 1.70 \\
4L-PbI$_2$&74.9  & 75.0  & 99.0  & 99.2  & 1.3&0.6
&5.8 &32.0 &5.5 & 4.28 & 1.89 \\
5L-PbI$_2$& 75.3  & 75.3 & 100.0 & 99.9 & 1.0&0.5
&9.0 &56.5 &6.3 & 4.68 & 2.06 \\
6L-PbI$_2$&75.7  & 75.5 &100.6  & 100.2 & 0.6&0.5
& 12.0& 83.7&7.0 & 4.81 & 2.18 \\
Bulk-PbI$_2$&     
75.8 & 75.8 & 101.1(96.0)\cite{Zhong} & 101.1 &$-$ &$-$
  & 32.3&384.4 &11.9 & 6.87 & 5.68 \\
\hline\hline 
\end{tabular}
\end{table*}

\begin{figure*}
\includegraphics[width=18cm]{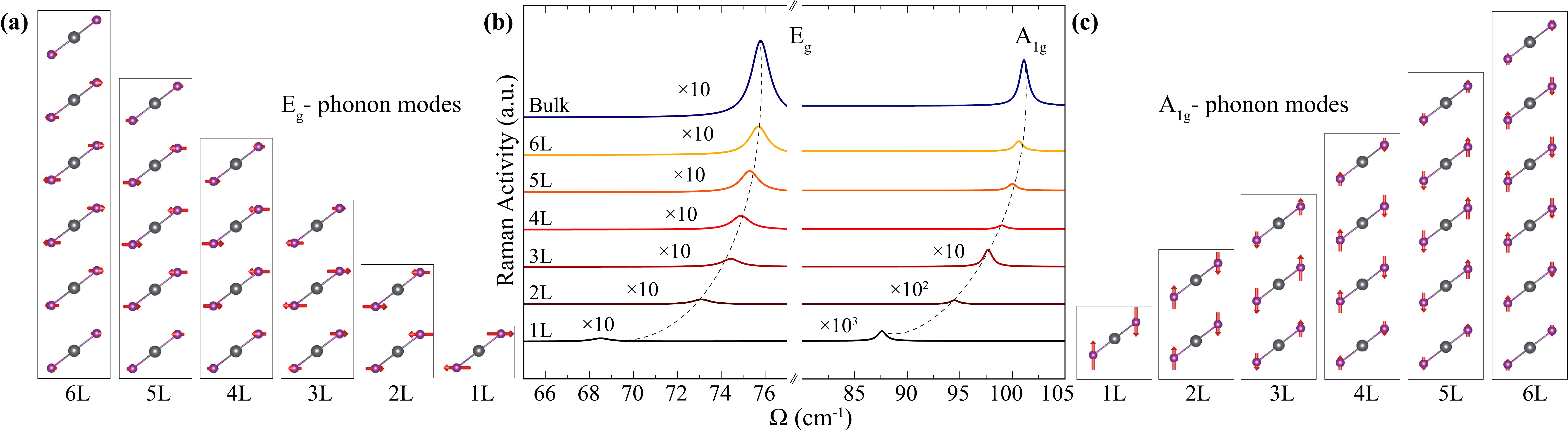}
\caption{\label{f5}
(color line) (a) The vibrational motion of I atoms in E$_{g}$ phonon modes for 
different number of PbI$_2$ layers. (b) The 
evolution of 
Raman spectrum with respect to the number of layers for the two characteristic 
prominent peaks. (c) The vibrational motion of I atoms in A$_{1g}$ phonon mode.}
\end{figure*}

\begin{figure}
\includegraphics[width=8.5cm]{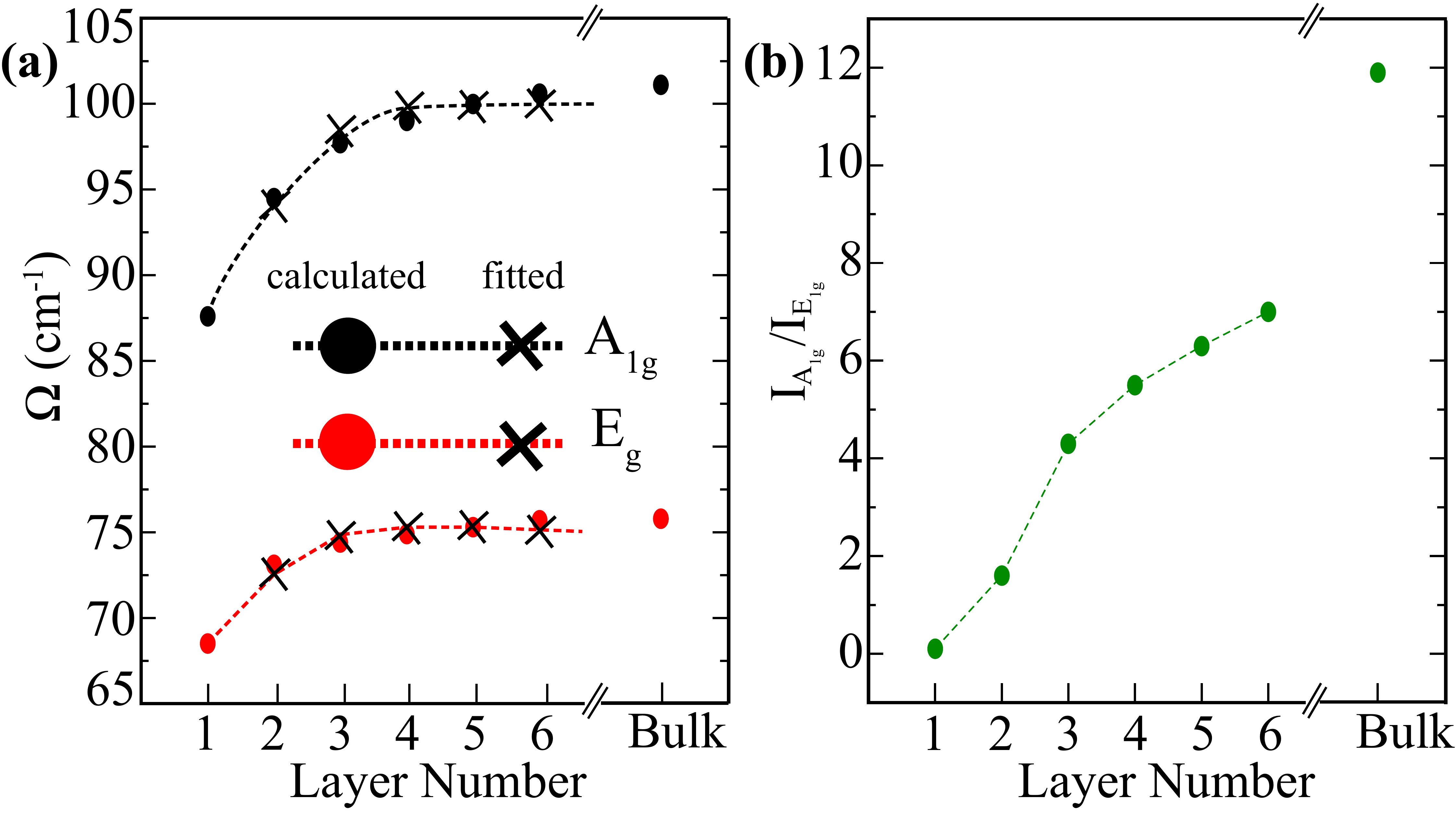}
\caption{\label{f6}
(color line) (a) The shifts in frequencies 
of E$_{g}$ and A$_{1g}$ vibrational modes with respect to the number of layers. 
(b) 
The 
change in Raman 
activity 
ratio of 
the two prominent peaks as a function of number of the layers. }
\end{figure}

\subsection{Phonons and Raman Spectrum}\label{vibrational}

In this section, we discuss the thickness dependency of the phononic properties 
of PbI$_2$, through high-frequency optical 
phonons and low-frequency layer breathing (LB), and inter-layer shear (C) modes 
by 
considering their 
frequencies and Raman activities. Note that, in crystals where vacuum is 
introduced, the Raman activities are normalized with respect to vacuum by using 
the thickness, $h$ (see Table \ref{main}), of the PbI$_2$ crystals. 

The dynamical stability of each PbI$_2$ crystal is examined by calculating the 
corresponding phonon band structure through whole BZ. As shown in Fig. 
\ref{f4}, all crystals are dynamically stable with no significant imaginary 
frequencies. Small negative frequencies
in the out-of-plane acoustic (ZA) mode near the $\Gamma$ point
are attributed to numerical artifacts which are caused by small inaccuracies of 
the FFT
grid. To determine the first-order off-resonant Raman spectrum, we calculate 
the 
zone-centered vibrational phonon 
modes at the 
$\Gamma$ point of the BZ. For a PbI$_2$ crystal there are two types of Raman 
modes, E$_g$ and A$_{1g}$. The E$_g$ modes are doubly degenerate and arise 
from the opposite in-plane vibration of
two I atoms with respect to the Pb atom while the A$_{1g}$ mode
is associated with the out-of-plane vibration of only I atoms in
opposite direction. Apart from those optical Raman 
modes, there are also low-frequency phonon modes which appear in the Raman 
spectrum in the low frequency region (generally below 50 cm$^{-1}$). These 
low-frequency modes
are categorized into two types: in-plane C and out-of-plane
LB vibrational modes. 

Initially, the majority of research activities in layered materials focused on
analyzing high-frequency optical phonon modes which involve vibrations of atoms 
that
stems from the intra-layer chemical bonds. These phonon modes which are called 
intra-layer modes, the
restoring forces are dominated by the strength of the intra-layer
chemical bonds rather than the vdW type forces which hold the layers together. 
Because of this reason, high-frequency intra-layer
modes are not very sensitive to the inter-layer coupling, and
therefore they are of limited use in the determination of thickness and
stacking order.

\subsubsection{\bf{High-frequency optical modes}}\label{hf}

As given in Table \ref{main2}, the 
peak 
frequencies of E$_g$ and A$_{1g}$ harden as the number of layers increases 
which 
is attributed to the inter-layer vdW forces suppressing 
atomic vibrations. The frequency of A$_{1g}$ mode displays an hardening from 
87.6 to 
101.1 cm$^{-1}$ when going from monolayer to bulk. In the case of 
E$_g$ 
mode, the corresponding frequency shifts from 68.5 to 75.8 cm$^{-1}$. 
The relative shift of both phonon modes, $\bigtriangleup\Omega_{A_{1g}}$ and 
$\bigtriangleup\Omega_{E_{g}}$, are calculated by using;

\begin{equation}\label{shiftrate} 
\bigtriangleup\Omega_i=\frac{\Omega_i(N)-\Omega_i(N-1)}{\Omega_i(N)},
\end{equation}and listed in Table \ref{main2}. As seen in Fig. \ref{f6}(a), 
as the number of layers increases the shift rate decreases and saturates to the 
bulk limit.

As in the case of the energy band gap, the evolution of the phonon 
frequencies with the number of layers can be fitted by the 
formula\cite{freqfit,freqfit2};

\begin{equation}\label{fit-freq} 
\Omega(N)=\Omega_{bulk}-D\frac{a}{N^{\gamma}},
\end{equation}where $\Omega_{bulk}$ is the frequency of the 
optical phonon mode for bulk crystal and $a$=4.45$\times$10$^{-8}$ cm is the 
lattice constant 
of bulk. 
$D$ and $\gamma$ are fitting parameters that match the $N$-dependent 
Raman shifts. For both of the prominent optical phonons, 
A$_{1g}$ and E$_g$, the calculated frequencies are fitted to Eq. 
(\ref{fit-freq}) and the fitted frequencies are listed in Table \ref{main2} and 
also are shown in Fig. \ref{f6}(a). When the calculated frequencies are fitted 
for A$_{1g}$, the parameters $D$ and $\gamma$ are found to be 
1.55$\times$10$^{-8}$ 
1/cm$^{2}$ and 1.83, 
respectively which gives the best fit to our calculated frequencies. In the 
case of E$_g$ phonon mode, $\gamma$=1.83 is found to be the same while 
$D$=0.64$\times$10$^{-8}$ 
1/cm$^{2}$ 
is smaller than that for A$_{1g}$. When fitting the calculated 
frequencies, the frequencies 
of the 1L-crystal are omitted because they do not exhibit the same trend of the 
few-layer structures. This is mainly attributed to the layer-layer interaction. 
As given in Table \ref{main2}, the frequency shift rates of both phonon modes 
are largest when going from 1L-crystal to the 2L-crystal. This is can be 
understood as follows: addition of a second layer induces additional springs 
between the 
layers that significantly increases the frequency. Our fitted function can be 
used to calculate the frequencies 
of both phonon modes for arbitrary thickness of PbI$_2$. Zhong \textit{et 
al.}\cite{Zhong} reported the frequency of A$_{1g}$ mode for 2L-, 9L-, and 
bulk-PbI$_2$ to be approximately equal ($\sim$96 cm$^{-1}$) while we find (see 
Table \ref{main2}) that they can differ by almost 5 cm$^{-1}$.

The suppression of atomic 
vibrations by the 
layer-layer vdW 
interaction is more dominant in bilayer and trilayer cases as supported by 
the values listed in 
Table \ref{main2}. As the number of layers increases, the relative contribution 
of the interaction with 
the outer neighboring  
layers decreases and thus, the change in the frequency gets 
smaller. The main contribution from the vdW interaction stems from the nearest 
neighboring layers in the center of the few-layer sample.

Zhang \textit{et al.}\cite{xzhang}, developed a diatomic chain model (DCM) for 
the 
intra-layer shear 
(E$_g$) 
and breathing (A$_{1g}$) modes that can explain the nature of the force 
constants 
per unit area, 
$\alpha^{\parallel}_{Pb-I}$ and $\alpha^{\perp}_{Pb-I}$, which are needed to 
describe the interaction between Pb and I atoms in a monolayer. Here, the 
component $\alpha^{\parallel}_{Pb-I}$ describes the in-plane lattice dynamics 
while the $\alpha^{\perp}_{Pb-I}$ determines that of the out-of-plane dynamics 
between Pb 
and 
I atoms. For these two optical phonon modes, the force constant per 
unit area can be related to the phonon frequency by the equations\cite{xzhang};
\begin{align}\label{optical} 
\Omega_{A_{1g}}=\left( 
\frac{1}{\sqrt{2}\pi c} \right)\sqrt{\frac{2\alpha^{\perp}_{Pb-I}}{\mu}}, 
\nonumber \\
\Omega_{E_g}=\left( 
\frac{1}{\sqrt{2}\pi c} \right)\sqrt{\frac{2\alpha^{\parallel}_{Pb-I}}{\mu}},
\end{align}
where $\mu$ is the atomic mass per unit area and $c$ is the speed of light. 
Due to the vibration of I atoms, the total mass per unit area is 
equal to 2m$_I$. Using the frequencies of A$_{1g}$ (87.6 cm$^{-1}$) and E$_g$ 
(68.5 cm$^{-1}$) in 1L-PbI$_2$ and the mass density of I atom 
(m$_I$=1.24$\times$10$^{-4}$ kg/m$^3$), we find the $\perp$ and $\parallel$ 
components of the force constant per unit area as; 
$\alpha^{\perp}_{Pb-I}$=0.34$\times$10$^{21}$ N/m$^3$ and 
$\alpha^{\parallel}_{Pb-I}$=0.21$\times$10$^{21}$ N/m$^3$ which are 
approximately 10 times smaller than those for MoS$_2$ (3.46$\times$10$^{21}$ 
and 1.88$\times$10$^{21}$ N/m$^3$, respectively)\cite{xzhang}. As listed in 
Table 
\ref{main4}, the $\alpha^{\parallel}_{Pb-I}$ is also much smaller than that of 
graphene (33.8$\times$10$^{21}$ N/m$^3$)\cite{g1} indicating the strong 
$C-C$ bonds 
in graphene that results in a high-frequency for the in-plane mode in 
graphene.

It has been shown for many other 2D layered materials that not only the peak 
frequencies but also the activities of Raman active 
modes are also key for the determination of the number of 
layers\cite{mos2,ws2,karl}. 
In 
the 
present study, we calculate the first order off-resonant 
Raman activities of two prominent, high-frequency optical phonons, E$_g$ and 
A$_{1g}$, for monolayer, few-layer, and bulk PbI$_2$ crystals. First of all, 
the individual Raman activities of each phonon mode display an increasing trend 
with increasing number of layers. Only in the monolayer 
limit, 
the Raman activity of A$_{1g}$ mode is much lower than that of E$_g$. In 
bilayer and few-layer cases, the contribution of both in-plane and out-of-plane 
dielectric constants to the Raman tensor increases. The reason why the 
increment 
in activity of A$_{1g}$ is much larger than that of E$_g$ can be explained 
through the Raman 
tensors of the two peaks. The Raman tensors of the two peaks are known from 
group symmetry of the crystal structure as:

\begin{equation}
R_{A_{1g}}=\left[\begin{array}{ccc}
a & 0 & 0 	\\
    0 & a & 
0 	\\
    0 & 0 & b 	\\
\end{array}\right] \\,
R_{E_{g}}=\left[\begin{array}{ccc}
c & 0 & 0 	\\
    0 & $-$c & d 	\\
    0 & d & 0 	\\
\end{array}\right],
\left[\begin{array}{ccc}
0 & $-$c & d 	\\
   $-$ c & 0 & 0 	\\
    d& 0 & 0 	\\
\end{array}\right] \nonumber
\end{equation}
where $a$, $b$, $c$, and $d$ are the derivative of the 
polarizability with respect to the considered normal mode. Since E$_g$ is 
doubly 
degenerate, the total 
Raman activity is the sum of the activities of two tensors standing for 
longitudinal and 
transverse orientations. In contrast to the Raman 
tensors of E$_g$ mode, there is an out-of-plane contribution of the derivative 
of polarizability 
in the Raman tensor of A$_{1g}$ (the number $b$). It can be clearly seen that 
increasing the in-plane dielectric constant from monolayer to bulk affects the 
Raman 
tensors of 
both modes (i.e. the values of R$_{11}$ and R$_{22}$ are affected). However, 
increasing the number of layers results in an increase of out-of-plane 
dielectric 
constant which only influences the Raman activity of the A$_{1g}$ mode. Another 
reason for the higher increase of Raman activity of  A$_{1g}$ is the 
contribution of the isotropic part of polarizability derivatives to the Raman 
activity. As given in 
Eq. (\ref{activity-final}), the isotropic part, $\tilde{\alpha}_s^2$, is the 
sum 
of 
the squares of the diagonal terms which is dominant 
in the anisotropic part, $\beta^2$. 

Also it should be noted that in Raman 
experiments a certain polarization direction is used to detect the Raman active 
phonon modes. As seen from the Raman tensors, the A$_{1g}$ mode is observable 
for only certain polarization angles while the E$_g$ mode is always observable 
independent of the polarization angle of the incident light. For a 
back-scattering configuration, the polarization vector of incident ($e_i$) and 
scattered ($e_s$) light are in the $xy$-plane. These two vectors can be 
represented 
in terms of an angle, $\theta$, which is the angle between the polarization 
vectors 
of incident and scattered light. Setting $e_i$ as ($\cos\theta$, $\sin\theta$, 
0) and $e_s$ as (1, 0, 0), one may calculate the Raman activities to be 
proportional to $a^2\cos^2\theta$ for A$_{1g}$ and proportional to $c^2$ for 
E$_g$ mode. Thus, the activity of E$_g$ mode is independent of the 
polarization angle $\theta$ while that of A$_{1g}$ is only non-zero when 
the polarization directions of incident and scattered lights are not 
perpendicular to each other. 

Since measured Raman intensities are taken on 
different substrates, the Raman intensities can also vary for different 
experimental setups (i.e. for different laser energies). Therefore, the 
discussion 
of relative Raman activities of 
the two prominent peaks seems to be more reliable for the determination of 
the number of layers in layered materials. In this section, we discuss the 
Raman 
activity ratio of A$_{1g}$ to 
that of E$_g$, i. e. $\frac{I_{A_{1g}}}{I_{E_g}}$. Our results reveal that in 
the 
monolayer limit, the Raman activity of A$_{1g}$ mode is lower than that of 
E$_g$ and the corresponding ratio is about 0.1 (see Fig. \ref{f6}(b)). Thus, 
the relative activity 
of A$_{1g}$ can be used to determine the thickness of a PbI$_2$ 
sample, i. e. the number of layers. 
As the numbers of layer increases, the Raman activity of A$_{1g}$ becomes 
dominant to that of E$_{g}$ and the ratio increases even in bilayer case. 

\begin{table*}
\caption{\label{main3} From 2L-to-6L PbI$_2$ crystal, the frequencies and Raman 
or infrared activity of C and LB modes.
Each color represents the same vibrational mode and assigned to the same 
description used in Figs. \ref{f7} and \ref{f8}. The following notations are 
used; R: 
Raman active, IR: Infrared active, and IR+R: both infrared and Raman active.}
\begin{tabular}{rcccccccccccccccc}
\hline\hline
& & &  & & & & 
  
\\
& C$_1$ & C$_2$ & C$_3$ & C$_4$ &C$_5$ 
&LB$_1$ 
&LB$_2$ & 
LB$_3$ 
&LB$_4$&LB$_5$\\
& (cm$^{-1}$) & 
(cm$^{-1}$) & (cm$^{-1}$) &(cm$^{-1}$) & (cm$^{-1}$) & (cm$^{-1}$) & 
(cm$^{-1}$) & (cm$^{-1}$)& 
(cm$^{-1}$) & 
(cm$^{-1}$) \\
\hline
2L-PbI$_2$&14.2  & $-$  &$-$ &$-$ 
&$-$ &24.5 &$-$ & $-$ & $-$ & $-$\\
& \textcolor{red}{(R)} &   & & 
& &\textcolor{red}{(R)} & &  &  & \\
3L-PbI$_2$& 17.4  & 10.0 &$-$ &$-$
&$-$ &16.8 &30.8 & $-$ & $-$ & $-$\\
&\textcolor{red}{(IR+R)}  & (R)  & & 
& &\textcolor{red}{(R)} &(IR) &  &  & \\
4L-PbI$_2$&18.6  & 14.2  & 7.5&$-$
&$-$ &12.7 &24.4& 33.4 &$-$ &$-$\\
& \textcolor{red}{(R)} & (IR)  &\textcolor{green}{(R)} & 
& & \textcolor{red}{(R)}&(IR) & \textcolor{green}{(R)} &  & \\
5L-PbI$_2$& 19.3  & 16.4  & 12.0 &6.2
&$-$ &10.3 &20.1& 28.6 &34.9 &$-$\\
& \textcolor{red}{(IR+R)} & (R)  &\textcolor{green}{(IR+R)} 
&\textcolor{blue}{(R)} 
& &\textcolor{red}{(R)} &(IR) & \textcolor{green}{(R)} & \textcolor{blue}{(IR)} 
&  \\
6L-PbI$_2$&19.7  & 17.5 & 14.3 &10.0
& 5.0& 8.7&17.0 & 24.7 & 31.1 & 36.1\\
& \textcolor{red}{(R)} &  (IR) & \textcolor{green}{(R)}& \textcolor{blue}{(IR)}
&\textcolor{cyan}{(R)} &\textcolor{red}{(R)} & (IR)& \textcolor{green}{(R)} & 
\textcolor{blue}{(IR)} &\textcolor{cyan}{(R)} \\
\hline\hline 
\end{tabular}
\end{table*}

\subsubsection{\bf{Interlayer shear and layer breathing modes}}\label{lf}

Zero-shift corresponds to Rayleigh (elastic) 
scattering of photons which has a very high intensity as compared to 
inelastically scattered photons. Since the inter-layer C and 
LB phonons 
have usually very low frequencies
(several to tens of wave numbers), the probing of these phonons through Raman 
spectroscopy is challenging. The
low-frequency characteristic of the inter-layer C and LB phonon modes actually
results from the weak inter-layer vdW restoring force. It was shown for 
other layered materials, such as graphene\cite{g1,g2} and MoS$_2$\cite{xzhang}, 
that these low-frequency phonon modes give information about the number of 
layers, 
$N$, 
since the vibrations
themselves are rigid motions of each layer. In contrast to the high-frequency 
optical phonons, the inter-layer 
modes have low frequencies and are almost completely determined by the
inter-layer restoring forces. The weak nature of the vdW layer-layer
interaction and the fact that a large ensemble of
atoms move together is responsible for the low frequencies which typically 
yields frequencies well below $\sim$100 
cm$^{-1}$. Due to their layer sensitivity to
inter-layer coupling, low-frequency Raman modes have recently started to
attract increasing attention for the determination of the
interfacial coupling and the thickness of the sample\cite{phtan}.

\begin{figure}
\includegraphics[width=8.5cm]{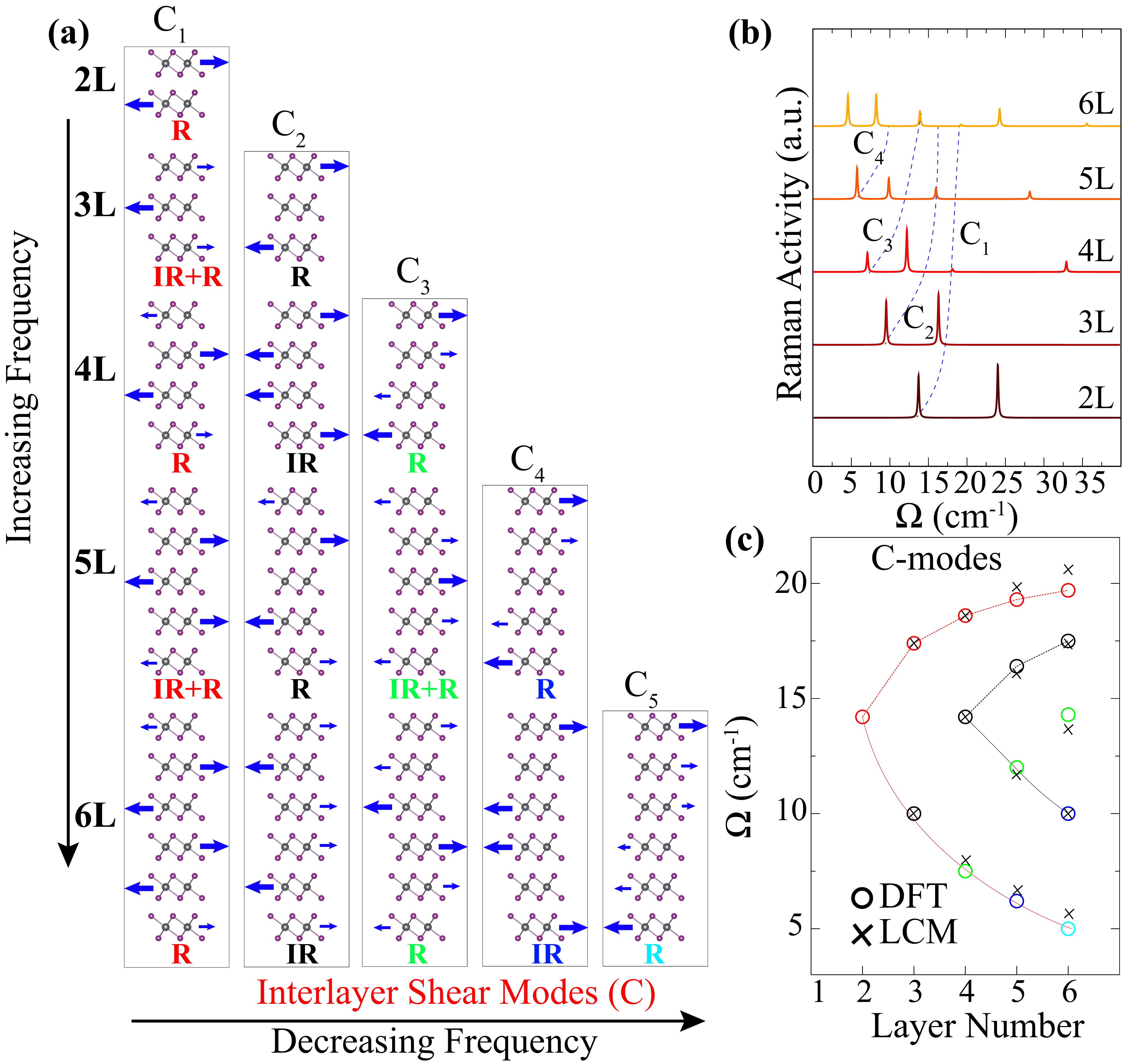}
\caption{\label{f7}
(color line) (a) The vibrational character of 
inter-layer C modes from 2L-to-6L. 
The Raman activity (R) and 
infrared activity (IR) of each phonon mode are given below the mode. The 
same colors correspond to the same phonon modes in different few-layer PbI$_2$. 
(b) The Raman spectrum of low-frequency
inter-layer C phonon modes for different number of layers of PbI$_2$ 
crystal. (c) The change in frequency of inter-layer C modes. Results of DFT 
calculations and LCM are compared. }
\end{figure}

\paragraph{\bf{Interlayer shear (C) modes:}}

The inter-layer C modes are assigned to the in-plane rigid-layer vibrations of 
each PbI$_2$ layer. The frequencies of C and LB modes 
are smaller than $\sim$50 cm$^{-1}$ which is a natural result of the weak vdW 
forces. 1T-PbI$_2$ belongs to the D$_{3d}$ point group which is independent of 
the 
number of 
layers. In contrast, the Raman or 
infrared 
activity of a C mode strongly depends on whether $N$ is even or odd.

For $N$-layer PbI$_2$ sample, one should count 2$\times$($N$-1) C 
modes 
where the coefficient 2 corresponds to the degeneracy 
of the modes. The C modes are either Raman or infrared active for even number 
PbI$_2$ (EN-PbI$_2$) layers while for odd 
number PbI$_2$ layers (ON-PbI$_2$), they are either Raman 
active or both infrared and Raman 
active. As seen on the right panel of Fig. \ref{f6}, one C 
mode appears in 2L-PbI$_2$. As the number of layers becomes 3, this mode splits 
into 
two branches one of which hardens and the other softens with increasing $N$. 
So for each number of layers, an additional mode appears
with increasing $N$. As seen in Fig. \ref{f6}(b) 
connecting each branch
of the C modes with dashed and solid lines shows a
series of cone-like curves. For example, the shear 
mode C$_1$ (denoted by red color) exhibits the opposite rigid vibration of 
each PbI$_2$ layer with respect to each other as shown in Fig. \ref{f6}. As $N$ 
increases 
from 2L-to-6L, its frequency hardens from 11.7 cm$^{-1}$ in 
2L-PbI$_2$ to 19.3 cm$^{-1}$ in 6L-PbI$_2$ and reaches 20.3 cm$^{-1}$ in 
bulk-PbI$_2$. 

As modeled for other 2D 
layered materials, the physics of C and LB modes can be
obtained using a simple linear chain model (LCM). Since 
each PbI$_2$ layer exhibits a rigid vibration, they can be 
considered as a single mass and then the LCM is constructed. Such approximation 
has
been proven to work very well for 2D layered materials\cite{g1,g2,xzhang}. The 
frequency of C$_1$ in bulk crystal is related to that of 2L-PbI$_2$ by the 
relation;

\begin{equation}\label{9} 
\Omega(C_{2,1})=\Omega(C_{bulk})/\sqrt{2}. 
\end{equation}
Using the relation given by Eq. (\ref{9}), the frequency of C$_1$ in the bulk 
limit 
is calculated to be 20.1 cm$^{-1}$ which is very close to that of bulk 
crystal (20.3 cm$^{-1}$) calculated within DFT. By the same methodology, one 
can 
calculate 
the frequencies of all C modes for bulk crystals by using the calculated C$_i$ 
values which are listed in Table \ref{main3}. 

As we relate the bulk frequency of a C mode to its frequency in $N$-PbI$_2$ by 
Eq. (\ref{9}), it is also possible to generate all the 
C mode frequencies from that of 2L-PbI$_2$ crystal. As stated 
by Zhang \textit{et al.}\cite{xzhang}, their LCM is applicable to any layered 
material. They reported that the general approach is
to calculate the $\mu$ for the monolayer of a given
material, and then from the knowledge of the frequency of C in 2L sample, one
can predict the relation between the frequency and $N$ for the
different branches in any layered material. The relation between the frequency 
of 
C modes with $N$ is given by the formula:

\begin{equation}\label{10} 
\Omega_C(N)=\Omega_C(2)\sqrt{1\pm cos\left(\frac{N_0\pi}{N}\right)},
\end{equation}
where $\Omega_C(N)$ is the frequency of the C mode in $N$-PbI$_2$ while 
$\Omega_C(2)$ represents that of the 2L sample and $N_0$ is an integer, 
$N_0$=1,2,3,4,... As listed in Table \ref{main3}, the frequency of C in 
2L-PbI$_2$ is found to be 14.3 cm$^{-1}$. Using Eq. (\ref{10}), one can find 
the 
frequencies of the two branches in 3L-PbI$_2$ 
which are 
17.5 and 10.1 cm$^{-1}$ for the higher and lower branches, respectively. These 
values are very close to the frequencies calculated directly by the small 
displacement method (17.4 and 10.0 cm$^{-1}$ 
for higher and lower branches, respectively). It is obvious that for the C 
modes in layered materials, the LCM matches well with the calculated 
frequencies using the small-displacement methodology.

As in the case of high-frequency optical modes, the inter-layer C mode 
frequency 
can also be represented in terms of the force 
constant per unit area, $\alpha$, and the reduced 
mass of a rigid layer, $\mu$, as:

\begin{equation}\label{c-bulk-frequency} 
\Omega_C=\left( 
\frac{1}{\sqrt{2}\pi c} \right)\sqrt{\frac{\alpha^{\parallel}_{I-I}}{\mu}},
\end{equation}
where $\alpha^{\parallel}_{I-I}$ denotes the in-plane nearest-neighboring
inter-layer force constant per unit area between two I atoms and $c$ is the 
speed of light. Because of the rigid vibration of each layer, one can assume 
one layer as a ball with mass m$_{Pb}$+2m$_I$. This relation allows us to 
calculate the force constant 
k$^{\parallel}_{I-I}$=$A\times\alpha^{\parallel}_{I-I}$ where $A$ is the area of 
the unit cell. The individual mass 
densities of Pb and I atoms per unit cell area are 
m$_{Pb}$=2.02$\times$10$^{-7}$ and m$_I$=1.24$\times$10$^{-4}$ kg/m$^3$, 
respectively. Now, using these mass densities in Eq. (\ref{c-bulk-frequency}) 
we 
find the inter-layer force constant per unit area, $\alpha^{\parallel}_{I-I}$, 
for the C mode in 2L-PbI$_2$ as $\alpha^{\parallel}_{I-I}$=1.61 
$\times$10$^{19}$ N/m$^3$ which is lower than that reported for MoS$_2$ 
(2.82 $\times$10$^{19}$ 
N/m$^3$)\cite{xzhang} (see Table \ref{main4}). This is exactly the reason why 
the 
frequency of C mode in 
2L-PbI$_2$ (14.3 cm$^{-1}$) is lower than that of 2L-MoS$_2$ (23.0 
cm$^{-1}$)\cite{xzhang}. It is also possible to calculate the force constant, 
k$^{\parallel}_{I-I}$, between 
two PbI$_2$ layer by multiplying $\alpha^{\parallel}_{I-I}$ by the unit cell 
area 
which gives 2.8 N/m which is slightly larger than that reported for 
MoS$_2$ (2.7 N/m)\cite{xzhang}. Moreover, the inter-layer shear modulus can 
also be calculated by multiplying $\alpha^{\parallel}_{I-I}$ by the 
equilibrium distance between
two adjacent PbI$_2$ layers which is the effective thickness of the monolayer 
crystal (7.13 \AA {}). The corresponding shear modulus is found to be 11.6 GPa 
which is 
lower than that of MoS$_2$ (18.9 GPa)\cite{xzhang}. 

\begin{figure}
\includegraphics[width=8.5cm]{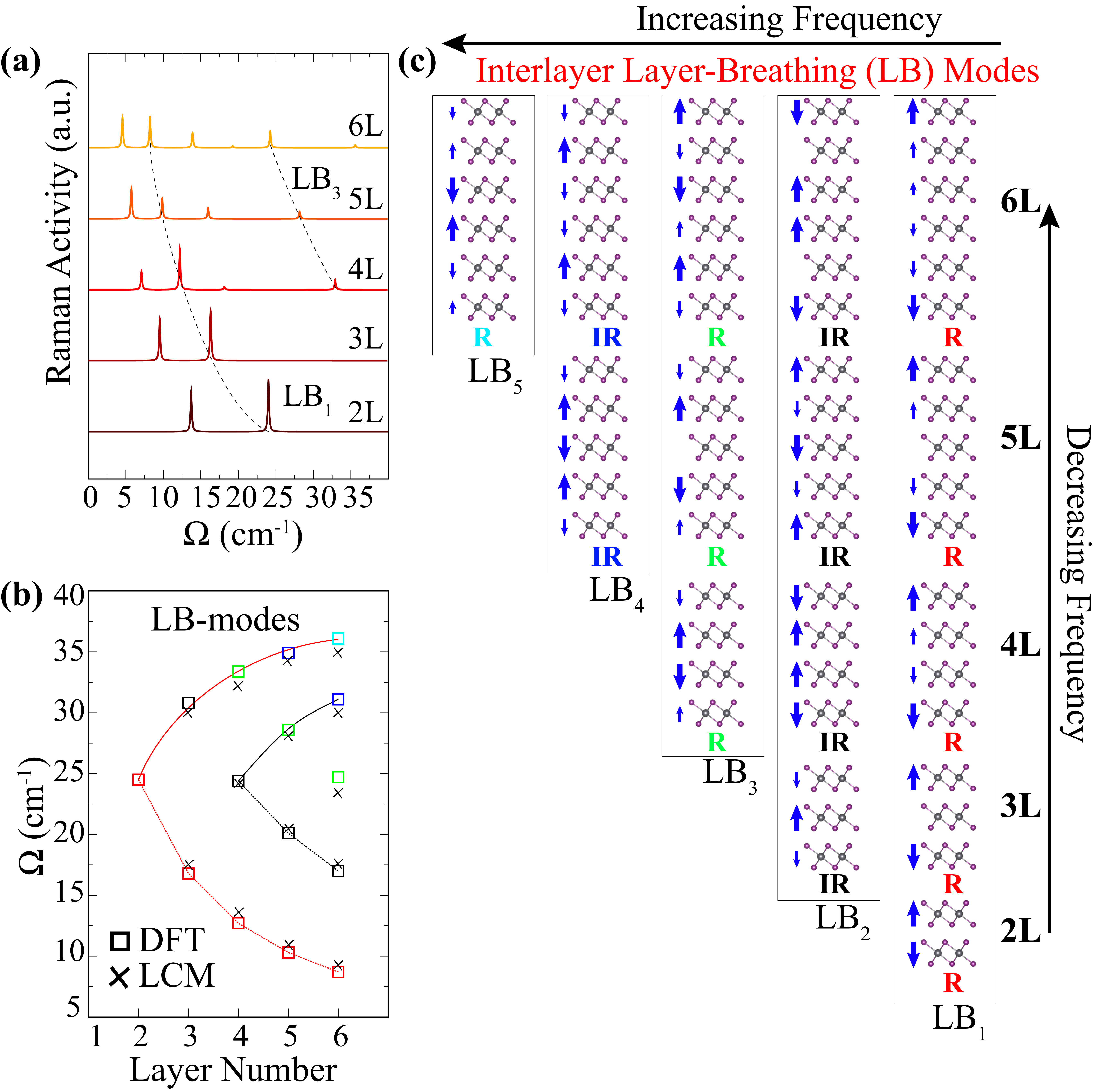}
\caption{\label{f8}
(color line) Left-panel; (a) the Raman spectrum of low-frequency
inter-layer breathing mode (LB) and the phonon modes for different number of 
layers of 
PbI$_2$ 
crystal. (b) The change in frequency of inter-layer LB modes 
with 
respect to number of layers. Results of DFT calculations and LCM are compared. 
(c) The vibrational character of 
inter-layer LB 
modes from 2L-to-6L. 
The Raman activity (R) and 
infrared activity (IR) of each phonon mode are given below the mode.}
\end{figure}

In addition to the peak frequencies, the Raman 
activity of the C modes strongly depends on the number of layers whether it is 
odd 
or even. The 
Raman activity of C$_1$ is distinguishable for EN-PbI$_2$ 
and ON-PbI$_2$ crystals. Our results reveal that the Raman activity 
values of C$_1$ in EN-PbI$_2$ are approximately 10$^4$ times that of 
ON-PbI$_2$. As listed in Table \ref{main3}, the C$_2$ modes (denoted by black 
color) are Raman active 
for ON-PbI$_2$ and infrared active for EN-PbI$_2$ and its frequency hardens 
from 9.2 to 17.2 cm$^{-1}$ from 3L-to-6L. The frequency evolution of the other 
C modes are also listed in Table \ref{main3}. As 
shown in the right-panel of Fig. \ref{f6}, for EN-PbI$_2$ crystals there are 
$\frac{N}{2}$ Raman active C modes while the remaining 
$\left(\frac{N}{2}-1\right)$ are infrared active. However, in the case of 
ON-PbI$_2$ the number of Raman active modes are $\left(\frac{N-1}{2}\right)$ 
and the remaining half of the C modes are both infrared and Raman active.

\paragraph{\bf{Inter-layer breathing (LB) modes:}}

In contrast to the C modes, the inter-layer LB modes are assigned to the 
out-of-plane rigid-layer vibrations of 
each PbI$_2$ layer. There are $N$-1 
non-degenerate LB modes in an $N$-PbI$_2$ crystal. Similar to the case of C 
modes, as $N$ increases each LB mode generates two branches one at higher and 
one at lower frequencies. The frequencies of the additional branches also obey 
the 
relation given in Eq. (\ref{10}). Moreover, the bulk frequency of any LB mode 
can 
also be predicted by using the relation given in 
Eq. (\ref{9}). 

\begin{table}
\caption{\label{main4} Parallel, $\alpha^{\parallel}_{Pb-I}$, and 
perpendicular, $\alpha^{\perp}_{Pb-I}$, force constants per unit area between 
Pb and I atoms. Those between I atoms from nearest neighboring PbI$_2$ 
layers, $\alpha^{\parallel}_{I-I}$, and $\alpha^{\perp}_{I-I}$. For comparison, 
the values for graphene and MoS$_2$ are also listed. The unit of the force 
constant 
per unit area is N/m$^{3}$.}
\begin{tabular}{rcccccccccccccccc}
\hline\hline
& $\alpha^{\parallel}_{Pb-I}$ &$\alpha^{\perp}_{Pb-I}$ & 
$\alpha^{\parallel}_{I-I}$ &$\alpha^{\perp}_{I-I}$\\
PbI$_2$&0.21$\times$10$^{21}$ & 0.34$\times$10$^{21}$ &1.61$\times$10$^{19}$ 
&4.78$\times$10$^{19}$ \\
\hline
& $\alpha^{\parallel}_{C-C}$ &$\alpha^{\perp}_{C-C}$ & 
$\alpha^{\parallel}_{C-C}$ &$\alpha^{\perp}_{C-C}$\\
Graphene\cite{g1}&33.8$\times$10$^{21}$  & $-$  & 
1.28$\times$10$^{19}$&10.7$\times$10$^{19}$\\
\hline
& $\alpha^{\parallel}_{Mo-S}$ &$\alpha^{\perp}_{Mo-S}$ & 
$\alpha^{\parallel}_{S-S}$ &$\alpha^{\perp}_{S-S}$\\
MoS$_2$\cite{xzhang}& 1.88$\times$10$^{21}$  & 3.46$\times$10$^{21}$ & 
2.82$\times$10$^{19}$ &8.90$\times$10$^{19}$\\
\hline\hline 
\end{tabular}
\end{table}

As in the case of the C modes, the total number of inter-layer LB modes depend 
oo the number of layers in the 
crystal. In an $N$-layer PbI$_2$, there exists ($N$-1) LB modes 
which are non-degenerate. The LB modes are either Raman active or infrared 
active depending on the number of layers in the PbI$_2$ crystal. As shown in 
the right-panel of Fig. \ref{f8}, the Raman active LB 
modes exist when the vibration is totally symmetric with respect to an axis 
perpendicular to the out-of-plane direction. For those LB modes, the 
out-of-plane 
vibration of layers has mirror symmetry along the out-of-plane direction. 
However, the infrared active LB modes do not exhibit such mirror 
symmetry that is why the dipole moment 
changes instead of the polarizability. As the number of layers increases, 
each LB branch generates additional 
branches one of which is Raman active and the other is infrared active. Thus, 
for EN-PbI$_2$ there occurs $\frac{N}{2}$ Raman active LB modes while the 
remaining 
$\left(\frac{N}{2}-1\right)$ are infrared active. In the case of 
ON-PbI$_2$ the number of Raman active modes is equal to the number of infrared 
active 
modes. By the same analogy with C 
modes, LB modes form a
series of cone-like curves as shown in Fig. \ref{f8}(c). For example, the 
LB$_1$ (denoted by red color) demonstrates the opposite rigid vibration of 
each PbI$_2$ layer with respect to each other in out-of-plane direction as 
shown in Fig. \ref{f8}(c). As $N$ 
increases 
from 2-to-6, its frequency softens from 24.5 cm$^{-1}$ in 
2L-PbI$_2$ to 8.7 cm$^{-1}$ in 6L-PbI$_2$. Moreover, the evolution of 
frequencies of LB modes with the number of layers $N$ can be explained by 
the 
relation given in Eq. (\ref{10}). For example, the LB mode of 2L-PbI$_2$ 
generates two additional branches in 3L-PbI$_2$ one with higher and the other 
with 
lower frequency. Using the frequency of 2L crystal we find the frequencies of 
the 
two branches in 3L-PbI$_2$ to 
be 30.0 and 17.3 cm$^{-1}$ for the higher and lower branches, respectively. 
These 
results agree with the frequencies calculated by using the 
small-displacement methodology. 

By using the relation given in Eq. (\ref{c-bulk-frequency}), one can calculate 
the out-of-plane nearest-neighboring
inter-layer force constant per unit area between two I atoms as, 
$\alpha^{\perp}_{I-I}$=4.78 
$\times$10$^{19}$ N/m$^3$ which is approximately half of that of MoS$_2$ 
(8.90 $\times$10$^{19}$ 
N/m$^3$)\cite{xzhang}. The corresponding inter-layer force constant is 
k$^{\perp}_{I-I}$=8.2 N/m which is slightly larger than the value for MoS$_2$ 
(7.8 
N/m)\cite{xzhang}. The difference between the force constant per unit area is 
therefore due to the larger unit cell area of PbI$_2$ when compared 
with that of MoS$_2$. The value for PbI$_2$ is also smaller than that of 
graphene as listed in Table \ref{main4}. The difference between different 
layered materials is due to the different inter-layer interactions between the 
individual 
layers. Thus, it is possible to conclude that the inter-layer interaction 
between PbI$_2$ layers in few-layer crystal is slightly smaller than those 
between graphene and MoS$_2$ layers. One also should note that as 
listed in Table \ref{main4}, the inter-layer force constants per unit area are 
approximately 100 times smaller than those for the intra-layer which means that 
in 
layered materials the intra-layer atomic bondings are much stronger than the 
inter-layer atomic interactions.

The LB mode in 2L-PbI$_2$ is found to be Raman active with a relatively high 
Raman activity as shown in Fig. \ref{f6}(a). As mentioned above, the 
generated branches harden with $N$ and are Raman active for EN-PbI$_2$ whose 
Raman activity display 
a decreasing trend. Thus, its observation 
becomes more difficult as $N$ increases. However, the soften one 
approximately conserves its Raman activity for different $N$ values. The reason 
for such different behavior in Raman activity can be understood through the 
strength of the vibrations of each layer. For the LB 
modes which soften with increasing number of layers, the vibration 
strength of the 
inner layers are much weaker than those of the outer layers. Apparently, 
the change of the polarizability and its volume is large. However, in 
EN-PbI$_2$ 
for the 
LB modes which harden as the number of layers increase, strong vibrations occur 
between the layers 
in the middle of the crystal and thus, the change of polarizability occurs in a 
relatively smaller volume which gives much smaller Raman activity. 
Although, the Raman activity changes from one LB mode to another and for 
different number of layers, the shift of the peak frequencies is more 
distinguishable for the determination of the layer-layer interaction and the 
number $N$ of layers 
rather than the Raman activities of the LB modes.

\section{Conclusions}\label{Conc}

In the present study, the number of layer-dependent electronic and vibrational 
properties of PbI$_2$ crystals were investigated by focusing on
the evolution of the band gap, peak frequencies, and corresponding 
activities of the Raman active phonon modes. Our results revealed that the 
direct or indirect gap semiconducting character of PbI$_2$ crystals are 
strongly 
influenced by 
the number of layers. In addition, an indirect-to-direct band gap transition is 
predicted for 3L-PbI$_2$. The layer-dependent Raman 
spectrum 
revealed that both prominent optical peaks, A$_{1g}$ and E$_{g}$, display 
phonon hardening with increasing number of layers which is attributed to the 
inter-layer vdW forces which suppress the atomic vibrations resulting in phonon 
hardening in directly stacked layered materials. Moreover, the relative Raman 
activities of A$_{1g}$ and E$_{g}$ peaks display an increasing trend from 
monolayer to bulk samples due to the strong enhancement of activity of A$_{1g}$ 
with increasing thickness which is especially important for the determination 
of the 
monolayer PbI$_2$. We further characterized rigid-layer vibrations both for
shear (C) and layer-breathing (LB) modes of few-layer PbI$_2$. Our study 
reveals 
that a reduced mono-atomic (linear) chain model (LCM) provides a
fairly accurate picture of the thickness dependence of the low-frequency
modes and is also a powerful tool to study the
inter-layer coupling strength in layered PbI$_2$.

\begin{acknowledgments}
 Computational resources were 
provided by TUBITAK ULAKBIM, High Performance and Grid Computing Center 
(TR-Grid 
e-Infrastructure). H.S. acknowledges financial support from the 
Scientific and Technological Research Council of Turkey (TUBITAK) under the 
project number 117F095. Part of this work was supported by FLAG-ERA project 
TRANS-2D-TMD.

\end{acknowledgments}

\end{document}